\documentclass{article}

\usepackage{amssymb,amsfonts,amsmath,stmaryrd}
\usepackage{enumerate,float,indentfirst}
\usepackage{color}
\usepackage{tikz}
\usetikzlibrary{arrows,snakes,backgrounds}
\usepackage[hidelinks,colorlinks=true,unicode]{hyperref}
\hypersetup{linkcolor=blue, citecolor=blue, filecolor=blue, urlcolor=blue}
\usepackage{url}
\usepackage[natbib=true, style=numeric-comp, sorting=none]{biblatex}
\def\be{\begin{eqnarray}}
\def\ee{\end{eqnarray}}
\def\nn{\nonumber}

\def\Tr{{\rm Tr}\,}

\definecolor{red}{rgb}{1,0,0}
\definecolor{orange}{rgb}{1,0.5,0}
\definecolor{violet}{rgb}{0.7,0,1}

\hoffset= - 1.0in         

\begin{document}

\hfill 

\hfill MIPT/TH-07/22

\hfill IITP/TH-08/22

\hfill ITEP/TH-10/22

\bigskip

\centerline{\Large{\bf{
Differential Expansion for antiparallel triple pretzels:
}}}
\centerline{\Large{\bf{
the way the factorization is deformed
}}}

\bigskip

\centerline{\bf{A. Morozov, N. Tselousov}}

\bigskip

\centerline{\it MIPT, ITEP \& IITP, Moscow, Russia}

\bigskip

\centerline{ABSTRACT}

\bigskip

{\footnotesize
For a peculiar family of double braid knots there is a remarkable
factorization formula for the coefficients of the  differential (cyclotomic) expansion (DE),
which nowadays is widely used to construct the exclusive
Racah matrices $S$ and $\bar S$ in arbitrary representations.
The origins of the factorization remain obscure and the
special role of double braids remains a mystery.
In an attempt to broaden the perspective,
we extend the family of double braids to antiparallel triple pretzels,
which are obtained by the defect-preserving deformation
from the trefoil and all have defect zero.
It turns out that factorization of DE coefficients is violated
quite strongly, still remains described by an elegant formula,
at least for all symmetric representations.

}

\bigskip

\bigskip

\section{Introduction}

An important task of modern theoretical physics is to study the hidden symmetries.
They are well known to play the crucial role in selecting of physically relevant
theoretical models.
The oldest example is provided by gauge symmetries \cite{Okun2},
which do not affect any physical observables,
but allow to give physical theories a simple and well distinguished formulation.
Another example is hidden integrability of exact non-perturbative (functional)
integrals (considered as functions of the coupling constants),
which are invariant under any change of integration variables,
which preserve the boundary conditions \cite{UFN3}.
The newly emerging concept, nicknamed {\it super}integrability \cite{MMsuperint1,MMsuperint2,MMsuperint3,IMMsuperint},
is the often-unexpected factorization of the answers for cleverly chosen
observables -- which turns out to reflect additional hidden properties
of the theory.
Such factorization is registered in many different circumstances,
and is not yet clearly explained in most cases --
often treated as  no more than  an amusing accident.
Still, the evidence is mounting that the phenomenon is quite general \cite{KM4,KM5}
and should have a universal explanation -- once again distinguishing between
the physically relevant models and artificial creations of the human mind.

In this paper we address a particular question about particular
factorization phenomenon -- that of the coefficients of the differential
expansion for HOMFLY-PT polynomials in knot  theory.
While looking rather special, this is quite a typical example --
it concerns the obviously-relevant $3d$ Chern-Simons theory,
and appropriate variables -- the coefficients of the {\it differential (cyclotomic) expansion}  \cite{IMMM,MMM2,AMM, MMM1, BM1,Morozov1, Morozov3,Morozov4, KNTZ1,CLZ,Morozov2, Morozov5, Morozov6, BJLMMMS, KM1,KM2,KM3, Itoyama:2012qt,Itoyama:2012re,Habiro2007,Nawata:2015wya,Chen:2014fpa,Chen:2015rid,Chen:2015sol,Kawagoe:2012bt,Kawagoe:2021onh,BG, Gorsky:2013jxa,Gukov:2011ry,Dunfield:2005si, berest2021, lovejoy2019colored, lovejoy2017colored, hikami2015torus,garoufalidis2005analytic,garoufalidis2011asymptotics} --
are far from obvious in the original formulation.
Still, it is slowly being recognized that these are (at least rather close to)
the right variables  \cite{AMM} and their properties are the ones that should be
studied by the future knot theory.
The first, still not truly impressive, attempts were made at the level of $\mathcal{C}$-polynomials
\cite{Cpols}.
A much more profound is the factorization property for the particular case
of double braid knots, discovered in \cite{Morozov6}
(see \cite{KNTZ1,Morozov1,Morozov3,Morozov4,CLZ} for numerous further developments).

This paper is an attempt to understand what is so special about double
braids, and what is the deformation of factorization property beyond them.
This can be an important step in the search for the truly relevant variables
beyond the double-braid example.
What we do, we consider the most straightforward generalization of
double braids -- to antiparallel triple pretzels.
This is the most general antiparallel evolution of the trefoil
and -- in accordance with the recent conjecture \cite{MT1}
about defect-invariance under antiparallel evolution --
all these pretzels have defect zero.
Thus this is indeed a kind of a {\it minimal} deformation of double braids
and the coefficients of DE remain defined in just the same way.
Still, as mentioned already in \cite{Morozov6}, factorization is dramatically violated.
We wonder what is the way to describe this violation.
The answer is an unexpectedly elegant formula -- much simpler than one
could anticipate and not immediately continued beyond the distingusihed
triple-pretzel family.
We do not know yet, what implications it will have,
but clearly it will help us to further advance the story of
"the proper variables" and factorization properties. 

As an additional test of our formulas for triple-pretzels we check that they possess 
still another factorization property -- 
of colored HOFMLY at the roots of unite \cite{KM4},
which is a far-going generalization of the celebrated factorization 
of special polynomials \cite{DMMSS,Zhu1}.
For even more amusing facts about HOMFLY at roots of unity see \cite{BMM}
and references therein.

\section{Differential expansion (DE)}

HOMFLY-PT polynomials are the Wilson-line averages in $3d$ Chern-Simons theory \cite{CS}:
\be
H_R^{\cal K} = \left<\Tr\!_R \, {\rm Pexp}\left(\oint_{\cal K} {\cal A}\right)\right>
\ee
which depend on integration contour (knot) ${\cal K}$,
the coupling constant $g$, the size $N$ of the $\mathfrak{sl}_N$ gauge algebra
and its representation $R$.
Dependence on $g$ and $N$ is actually through the
non-perturbative variables
$q=\exp\left(\frac{2\pi i}{g+N}\right)$ and $A=q^N$, of which $H^{\cal K}$
is -- mysteriously -- just a polynomial. For a recent progress in perturbative approach see \cite{Lanina:2021jzd,Lanina:2021nfj}.

Dependence on ${\cal K}$ and $R$ is more tricky,
and the better (less redundant) variables are the coefficients ${\cal F}_Q^{\cal K}$
of the {\it differential (cyclotomic) expansion} \cite{IMMM,MMM2,AMM, MMM1, BM1,Morozov1, Morozov2,Morozov3,Morozov4, Morozov5, Morozov6, BJLMMMS, KM1,KM2,KM3, Itoyama:2012qt,Itoyama:2012re,Habiro2007,Nawata:2015wya,KNTZ1,CLZ,Chen:2014fpa,Chen:2015rid,Chen:2015sol,Kawagoe:2012bt,Kawagoe:2021onh,BG, Gorsky:2013jxa,Gukov:2011ry,Dunfield:2005si, berest2021, lovejoy2019colored, lovejoy2017colored, hikami2015torus,garoufalidis2005analytic,garoufalidis2011asymptotics},
which we write down restricted in two ways --
to symmetric representations and to defect \cite{KM1} zero --
according to the limited consideration in this paper.
This allows to avoid lengthier formulas and extra notations,
unnecessary for what follows:
\be
H^{{\cal K}_0}_{[r]}(A,q) = 1
+ \sum_{s=1}^r \frac{[r]_q!}{[s]_q! [r-s]_q!}\cdot \underbrace{ {\cal F}_{[s]}^{{\cal K}_0}(A,q)}_{\text{depends on knot}}\cdot
\prod_{i=0}^{s-1} \{Aq^{r+i}\}\{Aq^{i-1}\}
\label{basicDE}
\ee
where $\{x\} := x - x^{-1}$ and $[n]_q := \frac{\{ q^{n} \}}{\{ q \}}$. We do not discuss origins of this formula and refer to other papers on the subject \cite{BM1, MT1}. The main advantage of DE formulas is that it extracts functions, such as ${\cal F}_Q^{\cal K}$, which truly distinguish knots from the other constituents of HOMFLY that are common for all knots. 
Equation (\ref{basicDE}) is literally true for defect-0 knots \cite{KM1}, 
thus ${\cal K}$ carries additional label $0$.
For other defects factorization is slightly less pronounced.
In this paper we deal with defect zero and do not need to go into more details on this difference.

The first example of zero defect knot is the trefoil knot $3_1$. In other words, zero defect means that the corresponding colored HOMFLY have the form \eqref{basicDE} with the following DE coefficients:
\be
\mathcal{F}^{3_1}_{[s]} = (-)^s A^{2s}q^{s(s-1)}
\ee

It is observed \cite{MT1} that the form of DE (i.e. the defect of DE) does not change under the antiparallel evolution of knots. By antiparallel evolution we mean replacement some of the crossings of knot by the antiparallel two-strand braid of odd length. In particular, this defect evolution hypothesis means that the whole three parametric family of triple antiparallel pretzels $\text{Pr}(2n-1,2m-1,2l-1)$ have the same form of DE \eqref{basicDE} as the trefoil knot, since they could be considered as descendants of the trefoil under antiparallel evolution \eqref{pic 3 pretz}. 

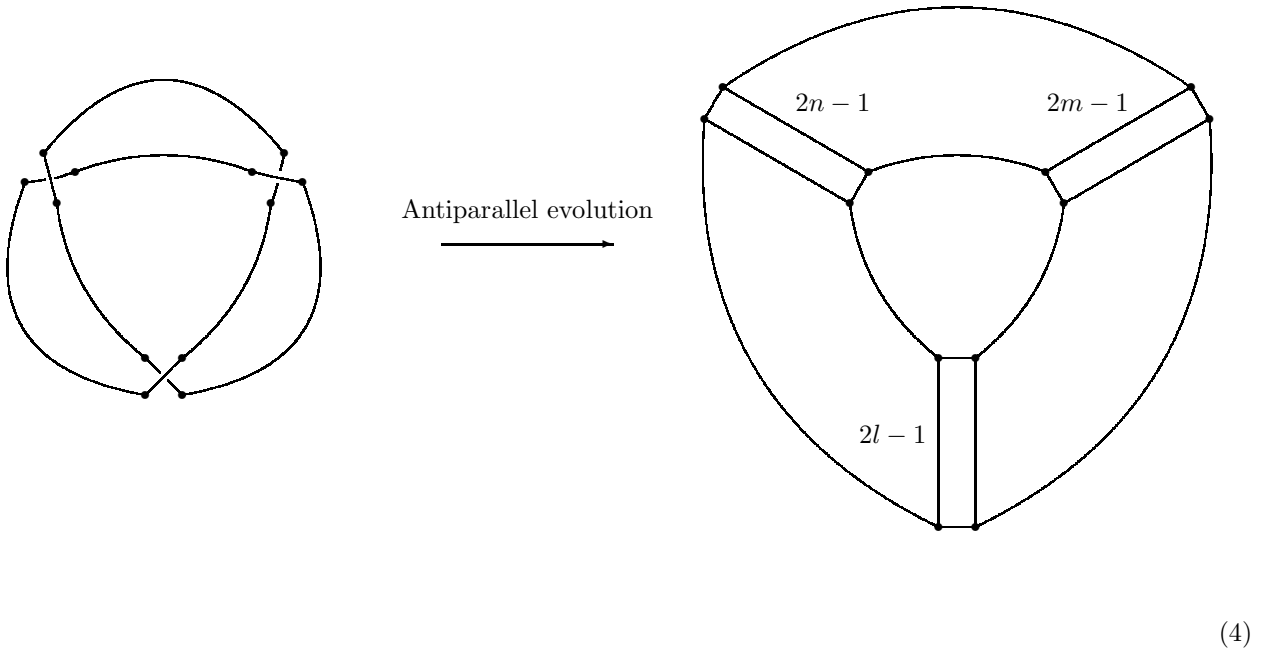
\be
\begin{picture}(400,250)(-40,-150)
\put(45.5,34.5){\circle*{3}}
\put(52.5,23.5){\circle*{3}}
\put(40.5,15.5){\circle*{3}}
\put(33.5,27.5){\circle*{3}}
\put(-45.6,34.5){\circle*{3}}
\put(-52.5,23.5){\circle*{3}}
\put(-40.5,15.5){\circle*{3}}
\put(-33.5,27.5){\circle*{3}}
\put(-7,-57){\circle*{3}}
\put(7,-57){\circle*{3}}
\put(-7,-43){\circle*{3}}
\put(7,-43){\circle*{3}}
\qbezier(45.5,34.5)(0,90)(-45.5,34.5)
\qbezier(33.5,27.5)(0,40)(-33.5,27.5)
\qbezier(52.5,23.5)(78,-45)(7,-57)
\qbezier(40.5,15.5)(36,-20)(7,-43)
\qbezier(-52.5,23.5)(-78,-45)(-7,-57)
\qbezier(-40.5,15.5)(-36,-20)(-7,-43)
\qbezier(52.5,23.5)(43,25)(33.5,27.5)
\qbezier(-40.5,15.5)(-43,25)(-45.5,34.5)
\qbezier(-7,-57)(0,-50)(7,-43)
\qbezier(45.5,34.5)(44.75,30.25)(44,28)
\qbezier(40.5,15.5)(41.5,17.25)(43,23)
\qbezier(-33.5,27.5)(-35,27)(-41,25)
\qbezier(-52.5,23.5)(-47,24)(-45,24.5)
\qbezier(-7,-43)(-4,-46)(-1.5,-48.5)
\qbezier(7,-57)(3,-53)(1.5,-51.5)
\put(105,0){\vector(1,0){65}}
\put(90,10){\mbox{Antiparallel evolution}}
%
%
\put(340.5,15.5){\circle*{3}}
\put(333.5,27.5){\circle*{3}}
\put(259.5,15.5){\circle*{3}}
\put(266.5,27.5){\circle*{3}}
\put(293,-43){\circle*{3}}
\put(307,-43){\circle*{3}}
\put(388.5,59.5){\circle*{3}}
\put(395.5,47.5){\circle*{3}}
\put(211.5,59.5){\circle*{3}}
\put(204.5,47.5){\circle*{3}}
\put(307,-107){\circle*{3}}
\put(293,-107){\circle*{3}}
\qbezier(333.5,27.5)(300,40)(266.5,27.5)
\qbezier(340.5,15.5)(336,-20)(307,-43)
\qbezier(259.5,15.5)(264,-20)(293,-43)
\qbezier(388.5,59.5)(300,120)(211.5,59.5)
\qbezier(395.5,47.5)(404,-60)(307,-107)
\qbezier(204.5,47.5)(196,-60)(293,-107)
\qbezier(307,-107)(300,-107)(293,-107)
\qbezier(293,-43)(300,-43)(307,-43)
\qbezier(307,-107)(307,-75)(307,-43)
\qbezier(293,-107)(293,-75)(293,-43)
\qbezier(340.5,15.5)(337,21.5)(333.5,27.5)
\qbezier(388.5,59.5)(392,53.5)(395.5,47.5)
\qbezier(333.5,27.5)(361,43.5)(388.5,59.5)
\qbezier(340.5,15.5)(368,31.5)(395.5,47.5)
\qbezier(259.5,15.5)(263,21.5)(266.5,27.5)
\qbezier(211.5,59.5)(208,53.5)(204.5,47.5)
\qbezier(266.5,27.5)(239,43.5)(211.5,59.5)
\qbezier(259.5,15.5)(232,31.5)(204.5,47.5)
\put(239,50.5){\mbox{$2n-1$}}
\put(334,50.5){\mbox{$2m-1$}}
\put(263,-75){\mbox{$2l-1$}}
\end{picture}
\label{pic 3 pretz}
\ee

The double braid knots are included in this family as $\text{Pr}(2n-1,2m-1,1)$ and posses surprisingly simple behaviour of DE coefficients:
\be
\label{double braid DE coef}
\boxed{
\mathcal{F}_{[s]}^{(n,m,1)} = \frac{F^{(n)}_{[s]} \cdot F^{(m)}_{[s]}}{F^{(1)}_{[s]}}
}
\ee
Here $(n,m,l)$ is a shortcut notation for $\text{Pr}(2n-1,2m-1,2l-1)$ and $F^{(n)}_{[s]}$ are the DE coefficients of twist knots:
\be
F^{(n)}_{[s]} := \mathcal{F}_{[s]}^{(n,1,1)}
\ee
because twist knots are included in triple pretzels as $\text{Pr}(2n-1,1,1)$. This mysterious factorization property of double braid DE coefficients \eqref{double braid DE coef} holds beyond symmetric and even rectangular representations \cite{Morozov4,Morozov3}. In order to understand the origins and nature of this factorization property we consider the double braid family as the part of a bigger family of triple pretzels and investigate how the factorization is deformed -- see the resulting formula for triple pretzels \eqref{3pretz result}.\\

The figure-eight knot $4_1$ ($n = -1$), trefoil $3_1$ ($n = 1$) and unknot $0_1$ ($n=0$) have simplest DE coefficients:
\begin{equation}
\label{simple DE coef}
    F^{(-1)}_{[s]} = 1, \hspace{15mm}
    F^{(0)}_{[s]} = 0, \hspace{15mm}
    F^{(1)}_{[s]} =  (-)^s A^{2s}q^{s(s-1)}
\end{equation}
for other twist knots there is a general formula \cite{MMM1}:
\be
F_{[s]}^{(n)} = q^{\frac{s(s-1)}{2}}A^s \cdot
\sum_{j=0}^s (-)^j \cdot \frac{[s]!}{[j]![s-j]!} \cdot \frac{(Aq^{j-1})^{2jn}\{Aq^{2j-1}\}}
{\prod_{i=j-1}^{s+j-1} \{Aq^i\}}
\label{twistDEcoef}
\ee
All these DE coefficients are encoded into the {\it one} semi-infinite lower triangular matrix $\mathcal{B}$ which we call the KNTZ matrix \cite{KNTZ1, Morozov1, Morozov4}:
\be
\label{F from KNTZ}
F^{(n)}_{[s]} = \sum_{j = 0}^{s} \left(\mathcal{B}^{n+1}\right)_{s,j}
\ee
where explicit form of matrix elements for $i,j = 0,1, 2, \ldots$:
\begin{align}
\label{KNTZ matr}
    {\cal B}_{i,j} &=(-)^{i-j} A^{2i}q^{(i+j)(i-1)}\frac{[i]_{q}!}{[j]_{q}! [i-j]_{q}!}, &\hspace{15mm} j \leqslant i \\
    {\cal B}_{i,j} &= 0, &\hspace{15mm} j > i
\end{align}
\begin{equation}
    {\cal B} = 
    \left( \begin{array}{ccccc}
         1 & 0 & 0 & 0 &\ldots \\
         -A^2 & A^2 & 0 & 0 & \ldots \\
         A^4 q^2 & -[2]_q A^4 q^3 & A^4 q^4 & 0 & \ldots \\
         -A^6 q^6 & [3]_q A^6 q^8 & -[3]_q A^6 q^{10} & A^6 q^{12} & \ldots \\
         \ldots & \ldots & \ldots & \ldots & \ldots \\
    \end{array}
    \right)
    \label{KNTZ}
\end{equation}
The approach to the twist knot DE coefficients through the KNTZ matrix turn out to be extremely productive: this structures generalizes to higher representations \cite{Morozov3} and superpolynomial T-deformation \cite{Morozov4}.

\section{Technicalities of HOMFLY-PT calculus for pretzel knots}

At present there is yet no direct way to calculate DE coefficients ${\cal F}_Q^{\cal K}$ for arbitrary knots and they are extracted from the answers for colored HOMFLY that are derived with the help of various techniques.  All pretzels are very simple examples of arborescent knots and
all the calculations are very straightforward due to the well-developed methods of arborescent calculus \cite{MMMRS, Chbili:2022pnt, Dhara:2017ukv, MMMRSS}. In this section we provide explicit formulas that allow to compute any symmetric HOMFLY for a 3-pretzel family that we discuss in this paper. 

The calculations involve only 4 matrices: $S, \bar S, T, \bar T$. The ${\cal R}$ matrix eigenvalues in symmetric representations $R = [r]$
in the two channels $[r]\otimes [r]$ and $[r] \otimes \overline{[r]}$ are very simple:
\be
T = {\rm diag} \Big((-)^{r + i - 1} q^{-r^2 + i(i - 1)} A^{-r} \Big),   \hspace{15mm}
\bar T = {\rm diag}\Big((-)^{r+i-1}q^{(i-1)(i-2)}A^{i-1}\Big),
\ee
for $i = 1, \ldots, r+1$.
For symmetric representations the matrices $S$,
which switch between $[r] \otimes \overline{[r]}$ and $\overline{[r]} \otimes \overline{[r]}$
and $\bar S$ acting within the space $\overline{[r]} \otimes \overline{[r]}$,
are also well known
\cite{LL3,MMS}:
for $i,j=1,\ldots, r+1$
\be
S_{ij} = \sqrt{\frac{\bar  d_i}{ d_j}}\cdot  \alpha_{i-1,j-1} \ \ \ \ \ \ \
\bar S_{ij} = \sqrt{\frac{\bar d_i}{\bar d_j}}\cdot \bar\alpha_{i-1,j-1}
\ee
where $d_X$ and $\bar  d_X$ are dimensions of representations from $[r] \otimes  [r]$
and $[r] \otimes \overline{[r]} $ respectively,
\begin{align}
\begin{aligned}
&[r] \otimes [r] = \bigoplus_{j = 1}^{r + 1} \ [r+j-1, r - j + 1]\\
d_j = {\rm dim}_{[r+j-1,r-j+1]} &:=\frac{[2j-1]_q}{\prod_{i=1}^{r+j}{[i]_q}\prod_{i=1}^{r+1-j} [i]_q}
\cdot \prod_{i=1}^{r+j-2}\{Aq^i\}\prod_{i=-1}^{r-1-j}\{Aq^i\}  \\
\end{aligned}
\end{align}
\begin{align}
\begin{aligned}
[r] \otimes \overline{[r]} =& \bigoplus_{i = 1}^{r + 1} \ [2(i-1), (i - 1)^{N-2} ]\\
\bar d_i = {\rm dim}_{[2(i-1), (i - 1)^{N-2}]}&:=\{Aq^{2i-3}\}\{Aq^{-1}\}\prod_{j=0}^{i-3} \left(\frac{\{Aq^j\}}{[j+2]}\right)^2
\end{aligned}
\end{align}
The matrices $S, \bar S$ are build from 6-j symbols \cite{}:
\be
\left(\begin{array}{c}
\alpha_{km}(r) \\ \\ \bar\alpha_{km}(r)
\end{array}\right)
= \frac{(-)^{r+k+m} [2m+1]\Big([k]![m]!\Big)^2[r-k]![r-m]!}{[r+k+1]![r+m+1]!}
\cdot \ \ \ \ \ \ \ \ \ \ \ \ \ \ \ \ \ \ \ \ \ \    \nn \\ \cdot
\sum_{j={\rm max}(r+m,r+k)}^{{\rm min}(r+k+m,2r)} \frac{(-)^j[j+1]!}{[2r-j]!\Big([j-r-k]! [j-r-m]![r+k+m-j]!\Big)^2}
\cdot\left(\begin{array}{c}
\frac{{\cal D}_{r-m}{\cal D}_{j+1}}{{\cal D}_{r+k+1}{\cal D}_{j-r-m} } \\  \\
\frac{{\cal D}_m^2{\cal D}_{j+1}}{{\cal D}_{r+k+1}{\cal D}_{r+m+1}{\cal D}_{r+k+m-j}}
\end{array}\right)
\label{alphasym}
\ee
where ${\cal D}_n:= \frac{1}{[n]!}\prod_{j=-1}^{n-2}\{Aq^j\}$
are responsible for the deviation from the $\mathfrak{sl}_2$ case when $A=q^2$ and ${\cal D}_n=1$.
The resulting answer for antiparallel 3-pretzel family:
\be
H_{[r]}^{(n,m,l)} = d_{[r]}^2 \cdot \sum_{j = 1}^{r + 1} d_j^{-1/2} \left(\bar S\bar T^{2n-1}S\right)_{1,j}
\left(\bar S\bar T^{2m-1}S\right)_{1,j}\left(\bar S\bar T^{2l-1}S\right)_{1,j}
\label{Hnml}
\ee
where $(n,m,l)$ is the shortcut notation for odd antiparallel triple pretzels $\text{Pr}(2n-1, 2m-1,2l-1)$. For example, $3_1 = (1,1,1)$ and $4_1 = (-1,1,1)$:
\begin{align}
        H_{[1]}^{(1,1,1)} = -A^4+A^2 q^2+\frac{A^2}{q^2}, \hspace{15mm} H_{[1]}^{(-1,1,1)} = A^2+\frac{1}{A^2}-q^2-\frac{1}{q^2}+1.
\end{align}

\section{Triple antiparallel pretzels}
In this section we analyze DE coefficents for triple antiparallel pretzels $\text{Pr}(2n-1,2m-1,2l-1)$.
All these knots have defect zero, thus the differential expansion is
\begin{align}
\begin{aligned}
H_{[1]}^{(n,m,l)} &= 1 + {\cal F}_{[1]}^{(n,m,l)}(A)\cdot \{Aq\}\{A/q\}  \\
H_{[2]}^{(n,m,l)} &= 1 + [2]\cdot {\cal F}_{[1]}^{(n,m,l)}(A)\cdot \{Aq^2\}\{A/q\}
+ {\cal F}_{[2]}^{(n,m,l)}(A,q)\cdot \{Aq^3\}\{Aq^2\}\{A\}\{A/q\}
\\
&\ldots  \\
H_{[r]}^{(n,m,l)} &= 1 + \sum_{s=1}^r \frac{[r]!}{[s]![r-s]!} \cdot {\cal F}_{[s]}^{(n,m,l)} \cdot
\prod_{j=0}^{s-1} \{Aq^{r+j}\}\{Aq^{j-1}\}
\end{aligned}
\end{align}

Dealing with triple pretzel DE coefficients we follow the strategy: we consider twist knot DE coefficients $F^{(n)}_{[s]}$ as "building blocks" and express the resulting answers through them, in the direct analogy with double braid family \eqref{double braid DE coef}. 

\subsection{The lowest coefficient ${\cal F}_{[1]}^{(n,m,l)}$}

The first DE coefficient is
 \be
{\cal F}_{[1]}^{(2n-1,2m-1,2l-1) } = -\frac{A^{2(n+m+l-1)}+A^{2(n+m+l-2)} - A^{2(n+m-1)}-A^{2(n+l-1)}-A^{2(m+l-1)} +1}{\{A\}^2}
\ee
so that the Alexander polynomial ($A = 1$ limit of HOMFLY polynomial) has the following form:
\be
{\rm Al}_{[1]}^{(2n-1,2m-1,2l-1)} = 1 - \underline{(nm+nl+ml-m-n-l+1)}\cdot\{q\}^2
\ee
and the underlined expression is independent of $q$, as required for defect-zero knots \cite{KM1,MT1}.

Remarkably
the first coefficient of the differential expansion is equal to
\be
\boxed{
{\cal F}_{[1]}^{(n,m,l)}
= \frac{F_{[1]}^{(n)}F_{[1]}^{(m)}F_{[1]}^{(l)}}{\left(F_{[1]}^{(1)}\right)^2}
+\frac{F_{[1]}^{(n-1)}F_{[1]}^{(m-1)}F_{[1]}^{(l-1)}}{F_{[1]}^{(1)}}
}
\label{F1 3-pretz}
\ee
i.e. has a simple expression through twist knot DE coefficients \eqref{twistDEcoef}:
\be
F_{[1]}^{(n)} = -\frac{A^2(A^{2n}-1)}{A^2-1} = A\cdot \frac{1-A^{2n}}{\{A\}}
\ee
Since $F_{[s]}^{(0)} = 0$, this formula reproduces factorization property \eqref{double braid DE coef} for all double braids with $(n,m,l)=(n,m,1)$. Also \eqref{F1 3-pretz} reproduces purely topological statement $(n,m,0) = (n - 1,m - 1, 1)$, due to $F^{(-1)}_{[s]} = 1$. 

\subsection{The second coefficient ${\cal F}_{[2]}^{(n,m,l)}$}

The second DE coefficient ${\cal F}_{[2]}^{(n,m,l)}$ has analogous first and second term as \eqref{F1 3-pretz} with an additional "correction":
\be
{\footnotesize
\boxed{
F_{[2]}^{(n,m,l)} =
 \frac{F_{[2]}^{(n)}F_{[2]}^{(m)}F_{[2]}^{(l)}}{\left(F_{[2]}^{(1)}\right)^2}
+\frac{F_{[2]}^{(n-1)}F_{[2]}^{(m-1)}F_{[2]}^{(l-1)}}{F_{[2]}^{(1)}}
-q^3[2]\left(F_{[2]}^{(n-1)}-F_{[1]}^{(n-1)}\right)
\left(F_{[2]}^{(m-1)}-F_{[1]}^{(m-1)}\right)\left(F_{[2]}^{(l-1)}-F_{[1]}^{(l-1)}\right)
}
}
\ee
with
\be
F_{[2]}^{(n)} = qA^2\cdot \left(\frac{1}{\{Aq\}\{A\}} - \frac{[2]A^{2n}}{\{Aq^2\}\{A\}}
+ \frac{q^{4n}A^{4n}}{\{Aq^2\}\{Aq\}}\right)
\ee

\subsection{Generic symmetric representation ${\cal F}_{[s]}^{(n,m,l)}$
\label{GSEtriple}}
Going through several examples we managed to write the following conjectural formula:
\begin{align}
\boxed{
\begin{aligned}
    \mathcal{F}_{[r]}^{(n,m,k)} &= \frac{F_{[r]}^{(n)}F_{[r]}^{(m)}F_{[r]}^{(k)}}{\left(F_{[r]}^{(1)}\right)^2}
+\frac{F_{[r]}^{(n-1)}F_{[r]}^{(m-1)}F_{[r]}^{(l-1)}}{F_{[r]}^{(1)}} + \\
& + \sum_{i = 1}^{r-1} (-)^{i} \left( A^2 q^{r-1} \right)^{2i} \cdot \frac{[r]!}{[i]! [r-i]!} \cdot \frac{G_{[r],i}^{(n)}\,G_{[r],i}^{(m)}\,G_{[r],i}^{(k)}}{F_{[r]}^{(1)}} \\
\end{aligned}
}\label{F_s 3-pretz}
\end{align}
where we define new functions $G_{[r],i}^{(n)}$ describing additional "correction" terms:
\be
\label{G functions}
G_{[r],i}^{(n)} := q^{i(r-1)} \cdot \sum\limits_{j=0}^{i} (-)^j q^{-j(i-1)} \cdot \frac{[i]_q!}{[j]_q! [i-j]_q!} \cdot F_{[r-j]}^{(n-1)}
\ee

{\it Surprisingly} the first and the second terms in \eqref{F_s 3-pretz} correspond to the "correction" summands for $i = 0$ and $i = r$:
\begin{align}
        &\frac{F_{[r]}^{(n-1)}F_{[r]}^{(m-1)}F_{[r]}^{(l-1)}}{F_{[r]}^{(1)}} = \frac{G_{[r],0}^{(n)}\,G_{[r],0}^{(m)}\,G_{[r],0}^{(k)}}{F_{[r]}^{(1)}} \\
    &\frac{F_{[r]}^{(n)}F_{[r]}^{(m)}F_{[r]}^{(k)}}{\left(F_{[r]}^{(1)}\right)^2} = (-)^{r} \left( A^{2r} q^{r(r-1)} \right)^{2} \cdot \frac{G_{[r],r}^{(n)}\,G_{[r],r}^{(m)}\,G_{[r],r}^{(k)}}{F_{[r]}^{(1)}}
\end{align}
due to $F_{[r]}^{(1)} = (-)^{r} A^{2r} q^{r(r-1)}$ and the following relations:
\begin{equation}
    G_{[r],0}^{(n-1)} = F^{(n-1)}_{[r]} 
\end{equation}
\begin{equation}
\label{mixed C pol}
    (-)^{r} \, G_{[r],r}^{(n)} = \frac{F^{(n)}_{[r]}}{F^{(1)}_{[r]}}
\end{equation}
The first relation follows directly from the definition \eqref{G functions}, while the second is less trivial and we discuss it in Sec. \ref{C pols sec}.
Finally, the resulting answer could be represented as extended tridiagonal sum:
\begin{align}
\label{3pretz result}
\boxed{\boxed{
\begin{aligned}
    \mathcal{F}_{[r]}^{(n,m,k)} &=  \sum_{i = 0}^{r} \, (-)^{i - r} \left( A^2 q^{r-1} \right)^{2i - r} \cdot \frac{[r]!}{[i]! [r-i]!} \cdot G_{[r],i}^{(n)} \cdot G_{[r],i}^{(m)} \cdot G_{[r],i}^{(k)} \\
\end{aligned}
}}
\end{align}

\section{Factorization at the roots of unity}
\subsection{Generalities}
It may seem that at $q=1$ the Chern-Simons theory becomes trivial, since quantization and all $\mathcal{R}$-matrices disappear.
However, this is not quite true for {\it reduced} HOMFLY polynomials,
because they are 
sensitive to the next order of the $1/N$ expansion,
according to the rule $q = \exp\left(\frac{2\pi i}{N+g}\right) = 1 + \frac{2\pi i}{N} + \ldots$. and $A=q^N$.
What happens, however, is that all the Wilson loop correlators in this approximation are factorized and {\it colored} HOMFLY are just powers of the fundamental ones
(which are sometimes called {\it special} polynomials \cite{DMMSS})
at  $q = 1:$
\be
\label{H at q = 1}
\left. H^{{\cal K}}_{R} \ \right|_{q = 1} = \left. \left(   H^{{\cal K}}_{[1]}  \right)^{|R|} \ \right|_{q = 1}
\ee
This implies some factorization property of the DE coefficients for symmetric HOMFLY. Namely,
substituting $q = 1$ to differential expansion for defect-zero knots \eqref{basicDE} and taking into account $ [s]_{q=1}  = s$ 
\be
\left. H^{{\cal K}}_{R} \ \right|_{q = 1} = 1
+ \sum_{s=1}^r \frac{r!}{s! (r-s)!} \cdot
\{ A \}^{2s} \cdot \left. {\cal F}_{[s]}^{\cal K} \right|_{q = 1}= \left( 1 + \{A\}^2 \cdot \left. {\cal F}_{[1]}^{\cal K}\right|_{q = 1}\right)^{r} = \left. \left(   H^{{\cal K}}_{[1]}  \right)^{r} \ \right|_{q = 1}
\ee
one concludes power law for DE coefficients of defect zero knots:
\be
\left. {\cal F}_{[s]}^{{\cal K}_{0}} \ \right|_{q = 1} =\left.  \left(  {\cal F}_{[1]}^{{\cal K}_{0}}  \right)^{s} \ \right|_{q = 1}
\label{q=1 DE coef}
\ee

Less trivially, similar factorizations occur at other unimodular values of $q= e^{\frac{ \pi \sqrt{-1}}{m}}$
\cite{KM4}:
\be
\label{factorization at q = i}
\left. H^{{\cal K}}_{[r]} \ \right|_{q = \sqrt{-1}} = \left.  H^{{\cal K}}_{[\text{Rem}(r / m)]}  \cdot \left(   H^{{\cal K}}_{[m]}  \right)^{\text{IntPart}(r / m)} \ \right|_{q = \sqrt{-1}}
\ee
where $r = m \cdot \text{IntPart}(r/m) + \text{Rem}(r/m)$. $\text{IntPart}$, $\text{Rem}$ are respectively the integer part and the reminder correspondingly. In this paper we analyze first nontrivial root of unity $q = \sqrt{-1}$ ($m=2$).
 
We can now check that these generic factorization properties are indeed respected
by our expressions for the triple pretzels \eqref{3pretz result}.

\subsection{$q = 1$}
 
Colored HOMFLY obey the special polynomial property \eqref{H at q = 1}
that is valid for arbitrary representation $R$ and arbitrary knot $\mathcal{K}$. This property is proved \cite{Zhu1} and may serve as a nontrivial check of our conjecture formula \eqref{3pretz result}.

To check this power law constraint for DE coefficients for our formula \eqref{3pretz result} we first compute it main building blocks $\left. F^{(n)}_{[r]} \right|_{q = 1}$ and $\left. G_{[r],i}^{(n)} \right|_{q = 1}$.
First, the twist knots are members of antiparallel 3-pretzel family and the corresponding DE coefficients ${\cal F}^{(n,1,1)}_{[r]} = F_{[r]}^{(n)}$ obey the power law constraint \eqref{q=1 DE coef}:
\be
\label{q = 1 F func}
\left. F^{(n)}_{[r]} \right|_{q = 1} = A^r \cdot \sum_{j=0}^{r} (-)^j \frac{r!}{j! (r-j)!} \cdot A^{2jn} \{ A \}^{-r} = \left( \frac{A}{\{ A \}} \left(A^{2n}-1\right)\right)^{r} =  \left. \left( F_{[1]}^{(n)}  \right)^{r} \right|_{q = 1}
\ee
Second, we use this result to evaluate \eqref{G functions} at the point $q = 1$:
\be
\left. G_{[r],i}^{(n)} \right|_{q = 1} = \sum_{j=0}^{i} (-)^{j} \frac{i!}{j! (i-j)!} \left. \left(F_{[1]}^{(n-1)}\right)^{r-j} \right|_{q = 1} = \left(F_{[1]}^{(n-1)}\right)^{r-i} \left. \left( F_{[1]}^{(n-1)} - 1\right)^{i} \right|_{q = 1}
\ee
and express in a form suitable for further analysis:
\be
\label{q = 1 G func}
\left. G_{[r],i}^{(n)} \right|_{q = 1} = \left. \left(G_{[1],0}^{(n)}\right)^{r-i} \left(G_{[1],1}^{(n)} \right)^{i} \right|_{q = 1}
\ee
where we used $\left. G_{[1],1}^{(n)} \right|_{q = 1} = \left.   F_{[1]}^{(n-1)} - 1  \right|_{q = 1}$ and $\left. G_{[1],0}^{(n)} \right|_{q = 1} = \left.   F_{[1]}^{(n-1)}  \right|_{q = 1}$. Applying relations \eqref{q = 1 F func} and \eqref{q = 1 G func} to 3-pretzel DE coefficients \eqref{3pretz result} at the point $q = 1$ we derive the power law behaviour \eqref{q=1 DE coef}:
\begin{align}
\begin{aligned}
\left. {\cal F}_{[r]}^{(n,m,l)} \right|_{q = 1} &= \left( - A^{-2} \cdot G_{[1],0}^{(n,m,l)} \right)^{r} \cdot \sum\limits_{i = 0}^{r} (-)^{i} A^{4i} \cdot \frac{r!}{i! (r-i)!} \cdot \left. \left(\frac{ G_{[1],1}^{(n,m,l)}}{G_{[1],0}^{(n,m,l)}} \right)^{i}\right|_{q = 1} = \\
& = \left. \left(- A^{-2} \cdot G_{[1],0}^{(n,m,l)} + A^2 \cdot G_{[1],1}^{(n,m,l)}\right)^{r} \right|_{q = 1}=\left. \left( {\cal F}_{[1]}^{(n,m,l)} \right)^{r} \right|_{q = 1}
\end{aligned}
\end{align}
where we define $G_{[r],i}^{(n,m,l)} := G_{[r],i}^{(n)} \cdot G_{[r],i}^{(m)} \cdot G_{[r],i}^{(l)}$.
\subsection{$q = \sqrt{-1}$}
\subsubsection{$q$-binomials at root of unity}
This case is less trivial as the case of $q = 1$ since the $q$-binomial coefficients evaluate to the following expression \cite{qBin}:
\be
\label{q-binom at i}
\left. \frac{[M]_q!}{[M-N]_q! [N]_q !} \right|_{q = \sqrt{-1}} = (-)^{ \frac{1}{2}  \text{Rem}(\frac{M+N-2}{4})  \cdot \text{Rem}(\frac{N}{4})} \cdot \frac{m!}{n! (m-n)!} \cdot \binom{r_m}{r_n}
\ee
where $M = 2 \cdot m + r_m$, $N = 2 \cdot n + r_n$. In other words, $m := \text{IntPart}(M/2), n := \text{IntPart}(N/2)$ and $r_m := \text{Rem}(M/2), r_n := \text{Rem}(N/2)$. The symbol is defined $\binom{r_m}{r_n} := \frac{r_m!}{r_n! (r_m - r_n)!}$ for $r_m \geqslant r_n$ and  $\binom{r_m}{r_n} := 0$ for $r_m < r_n$. We provide the Pascal triangle of $q$-binomial coefficients at the point $q = \sqrt{-1}$:
\begin{equation}
\setcounter{MaxMatrixCols}{15}
\begin{matrix}
&&&&& 1\\
&&&& 1 && 1\\
&&& 1 && 0 && 1\\
&& 1 && -1 && -1 && 1\\
& 1 && 0 && 2 && 0 && 1\\
1 && 1 && -2 && -2 && 1 && 1\\
\end{matrix}
\end{equation}
The "differentials" of the DE evaluate:
\begin{align}
   \left. \{ A q^{n} \} \right|_{q = \sqrt{-1}} = \left(\sqrt{-1}\right)^n \left(A - (-)^n A^{-1} \right) =  \left(\sqrt{-1}\right)^n  \cdot \begin{cases}
   \{ A \}, \hspace{15mm}  \text{even} \ n; \\
   \{ A \}_{+}, \hspace{15mm}  \text{odd} \ n; \\
   \end{cases}
\end{align}
where $\{A\}_{+} := A + A^{-1}$.
\subsubsection{Simple examples}
The factorization property can be seen from the simple examples:
\begin{equation}
    \left. H_{[1]}^{\mathcal{K}_0} \right|_{q = \sqrt{-1}} = 1 + \left. \mathcal{F}_{[1]}^{\mathcal{K}_0}  \cdot  \{A\}^{2}_{+} \right|_{q = \sqrt{-1}}
\end{equation}
\begin{equation}
    \left. H_{[2]}^{\mathcal{K}_0} \right|_{q = \sqrt{-1}} = 1 + \left. \mathcal{F}_{[2]}^{\mathcal{K}_0}  \cdot \{A\}^2 \{A\}^{2}_{+} \right|_{q = \sqrt{-1}}
\end{equation}
For representation $[r] = [3]$:
\begin{align}
\begin{aligned}
    \left. H_{[3]}^{\mathcal{K}_0} \right|_{q = \sqrt{-1}} &= 1 + 
    \left. \mathcal{F}_{[1]}^{\mathcal{K}_0} \cdot \{A\}^{2}_{+} + \mathcal{F}_{[2]}^{\mathcal{K}_0} \cdot \{A\}^2 \{A\}^2_{+} + \mathcal{F}_{[3]}^{\mathcal{K}_0}  \cdot \{A\}^2 \{A\}^4_{+} \right|_{q = \sqrt{-1}} \\
    \left. H_{[3]}^{\mathcal{K}_0} \right|_{q = \sqrt{-1}} &= \left. H_{[1]}^{\mathcal{K}_0} \cdot H_{[2]}^{\mathcal{K}_0} \right|_{q = \sqrt{-1}} \hspace{5mm} \Longrightarrow  \hspace{5mm} \left. {\cal F}_{[3]}^{\mathcal{K}_0} \right|_{q = \sqrt{-1}} = \left. {\cal F}_{[1]}^{\mathcal{K}_0} \cdot {\cal F}_{[2]}^{\mathcal{K}_0} \right|_{q = \sqrt{-1}}
\end{aligned}
\end{align}
For representation $[r] = [4]$:
\begin{align}
\begin{aligned}
    \left. H_{[4]}^{\mathcal{K}_0} \right|_{q = \sqrt{-1}} &= 1 + 
    \left. \mathcal{F}_{[2]}^{\mathcal{K}_0} \cdot \{A\}^2 \{A\}^2_{+} + \mathcal{F}_{[4]}^{\mathcal{K}_0}  \cdot \{A\}^4 \{A\}^4_{+} \right|_{q = \sqrt{-1}} \\
    \left. H_{[4]}^{\mathcal{K}_0} \right|_{q = \sqrt{-1}} &= \left. H_{[2]}^{\mathcal{K}_0} \cdot H_{[2]}^{\mathcal{K}_0} \right|_{q = \sqrt{-1}} \hspace{5mm} \Longrightarrow  \hspace{5mm} \left. {\cal F}_{[4]}^{\mathcal{K}_0} \right|_{q = \sqrt{-1}} = \left. {\cal F}_{[2]}^{\mathcal{K}_0} \cdot {\cal F}_{[2]}^{\mathcal{K}_0} \right|_{q = \sqrt{-1}}
\end{aligned}
\end{align}
\subsubsection{Arbitrary representations}
For arbitrary representation of even length $[r] = [2n]$:
\begin{equation}
    \left. H_{[2n]}^{{\cal K}_0} \right|_{q = \sqrt{-1}} = \sum_{s = 0}^{n} \frac{n!}{s!(n-s)!} \cdot \{ A \}^{2s} \{ A \}_{+}^{2s} \cdot \left. {\cal F}_{[2s]}^{{\cal K}_0} \right|_{q = \sqrt{-1}}
\end{equation}
where all the $q$-binomial coefficient are evaluated at the point $q = \sqrt{-1}$ following \eqref{q-binom at i}. From the factorization property \eqref{factorization at q = i} and the last formula follows the corresponding factorization of the DE coefficients:
\begin{equation}
    \left. H_{[2n]}^{\mathcal{K}_0} \right|_{q = \sqrt{-1}} = \left. \left( H_{[2]}^{\mathcal{K}_0} \right)^{n} \right|_{q = \sqrt{-1}} \hspace{5mm} \Longrightarrow  \hspace{5mm} \left. {\cal F}_{[2n]}^{\mathcal{K}_0} \right|_{q = \sqrt{-1}} = \left. \left( {\cal F}_{[2]}^{\mathcal{K}_0} \right)^n \right|_{q = \sqrt{-1}}
\end{equation}
For representations of odd length $[r] = [2n+1]$ the situation is a little more involved:
\begin{equation}
     \left. H_{[2n+1]}^{{\cal K}_0} \right|_{q = \sqrt{-1}} = \sum_{s = 0}^{n} \frac{n!}{s!(n-s)!} \cdot \{ A \}^{2s} \{ A \}_{+}^{2s} \cdot \left.\left(  {\cal F}_{[2s]}^{{\cal K}_0} + \{ A \}_{+}^{2}  {\cal F}_{[2s+1]}^{{\cal K}_0} \right) \right|_{q = \sqrt{-1}}
\end{equation}
when evaluation we used the following relations that appear from \eqref{q-binom at i}: $(-)^{\frac{1}{2} \text{Rem}\left( \frac{2n+2s-1}{4} \right) \cdot \text{Rem}\left( \frac{2s}{4} \right) } = (-)^{s}$ and $(-)^{\frac{1}{2} \text{Rem}\left( \frac{2n+2s}{4} \right) \cdot \text{Rem}\left( \frac{2s+1}{4} \right) } = (-)^{s+n}$. Therefore we conclude the factorization property for DE coefficients for zero defect knots:
\begin{equation}
    \left. H_{[2n+1]}^{\mathcal{K}_0} \right|_{q = \sqrt{-1}} = \left. H_{[1]}^{\mathcal{K}_0} \cdot  \left( H_{[2]}^{\mathcal{K}_0} \right)^{n} \right|_{q = \sqrt{-1}} \hspace{5mm} \Longrightarrow  \hspace{5mm} \left. {\cal F}_{[2n+1]}^{\mathcal{K}_0} \right|_{q = \sqrt{-1}} = \left. {\cal F}_{[1]}^{\mathcal{K}_0} \cdot \left( {\cal F}_{[2]}^{\mathcal{K}_0} \right)^n \right|_{q = \sqrt{-1}}
\end{equation}
\subsubsection{Direct check of \eqref{3pretz result}}
Now we evaluate main "building blocks" $\left. F^{(n)}_{[r]} \right|_{q = \sqrt{-1}}$ and $\left. G_{[r],i}^{(n)} \right|_{q = \sqrt{-1}}$ at the point $q = \sqrt{-1}$. From the explicit form \eqref{twistDEcoef} and \eqref{q-binom at i}:
\begin{align}
    \begin{aligned}
        \left. F_{[2s]}^{(n)} \right|_{q = \sqrt{-1}} = A^{2s} \cdot \sum_{j = 0}^{s} (-)^{j} \cdot \frac{s!}{j! (s - j)!} \cdot \frac{A^{4jn}}{\{A\}^s \{A \}_{+}^{s}} = \left( \frac{A^2}{\{A\} \{A\}_{+}}\right)^s \left( 1 - A^{4n} \right)^s = \left. \left( F_{[2]}^{(n)} \right)^s \right|_{q = \sqrt{-1}}
    \end{aligned}
\end{align}
For the odd length of the representations $[2s+1]$ the calculation involve splitting the sum for even and odd indices:
\begin{align}
\begin{aligned}
    \left. F_{[2s+1]}^{(n)} \right|_{q = \sqrt{-1}} &= A^{2s+1} \cdot \sum_{j = 0}^{s} (-)^{j} \cdot \frac{s!}{j! (s - j)!} \cdot \left( \frac{A^{4jn}}{\{A\}^{s+1} \{A \}_{+}^{s}} - \frac{A^{4jn + 2 n}}{\{A\}^{s+1} \{A \}_{+}^{s}} \right) = \\
    &= \frac{A}{\{ A \}} \left( 1 - A^{2n} \right) \cdot \left( \frac{A^2}{\{A\} \{A\}_{+}}\right)^s \left( 1 - A^{4n} \right)^s = \left. F_{[1]}^{(n)} \cdot \left( F_{[2]}^{(n)} \right)^s \right|_{q = \sqrt{-1}}
\end{aligned}
\end{align}
\begin{align}
    \begin{aligned}
        \left. G_{[2r],2i}^{(n)} \right|_{q = \sqrt{-1}} &= \left. \left( G_{[2],0}^{(n)} \right)^{r-i} \cdot \left( G_{[2],2}^{(n)} \right)^{i} \right|_{q = \sqrt{-1}} \\ 
        \left. G_{[2r+1],2i}^{(n)} \right|_{q = \sqrt{-1}} &= \left. (-)^{i} \cdot G_{[1],0}^{(n)}  \cdot G_{[2r],2i}^{(n)} \right|_{q = \sqrt{-1}} \\ 
        \left. G_{[2r+2],2i+1}^{(n)} \right|_{q = \sqrt{-1}} &= \left. (-)^{r} \cdot G_{[2],1}^{(n)}  \cdot G_{[2r],2i}^{(n)} \right|_{q = \sqrt{-1}} \\ 
        \left. G_{[2r+1],2i+1}^{(n)} \right|_{q = \sqrt{-1}} &= \left. (-)^{i+r} \cdot G_{[1],1}^{(n)}  \cdot G_{[2r],2i}^{(n)} \right|_{q = \sqrt{-1}} \\ 
    \end{aligned}
\end{align}
We check \eqref{factorization at q = i} for 3-pretzel DE coefficients \eqref{3pretz result} for even length of representations $[2r]$:
\begin{align}
    \begin{aligned}
        \left. {\cal F}^{(n,m,l)}_{[2r]} \right|_{q = \sqrt{-1}} &=\left( - A^{-4} \cdot G^{(n,m,l)}_{[2],0} \right)^{r} \cdot \sum_{i = 0}^{r} \ \frac{r!}{i! (r-i)!} \cdot A^{8i} \cdot \left. \left( \frac{G^{(n,m,l)}_{[2],2}}{G^{(n,m,l)}_{[2],0}} \right)^{i} \right|_{q = \sqrt{-1}} =\\
        &= \left. \left( - A^{-4}\cdot G^{(n,m,l)}_{[2],0}  - A^4 \cdot G^{(n,m,l)}_{[2],2} \right)^r \right|_{q = \sqrt{-1}} = \left. \left( {\cal F}_{[2]}^{(n,m,l)} \right)^r \right|_{q = \sqrt{-1}}
    \end{aligned}
\end{align}
and odd length of representations $[2r+1]$:
\begin{align}
    \begin{aligned}
        \left. {\cal F}^{(n,m,l)}_{[2r+1]} \right|_{q = \sqrt{-1}} &= \left( - A^{-2}  G^{(n,m,l)}_{[1],0} + A^{2}  G^{(n,m,l)}_{[1],1}  \right)  \left( - A^{-4}  G^{(n,m,l)}_{[2],0} \right)^{r}  \sum_{i = 0}^{r} \ \frac{r!}{i! (r-i)!} \cdot A^{8i} \cdot \left. \left( \frac{G^{(n,m,l)}_{[2],2}}{G^{(n,m,l)}_{[2],0}} \right)^{i} \right|_{q = \sqrt{-1}} =\\
        &= \left.  \left( - A^{-2} \cdot G^{(n,m,l)}_{[1],0} + A^{2} \cdot  G^{(n,m,l)}_{[1],1}  \right)  \left( - A^{-4}\cdot G^{(n,m,l)}_{[2],0}  - A^4 \cdot G^{(n,m,l)}_{[2],2} \right)^r \right|_{q = \sqrt{-1}} =\\
        &= \left. \left( {\cal F}_{[1]}^{(n,m,l)} \right) \cdot \left( {\cal F}_{[2]}^{(n,m,l)} \right)^r \right|_{q = \sqrt{-1}}
    \end{aligned}
\end{align}
where we define $G_{[r],i}^{(n,m,l)} := G_{[r],i}^{(n)} \cdot G_{[r],i}^{(m)} \cdot G_{[r],i}^{(l)}$.

\section{New ${\cal C}$-polynomial for the twist knots}
\label{C pols sec}
The colored HOMFLY polynomials are observed to satisfy linear recurrence relations with respect to representation \cite{Garoufalidis:2016zhf, Garoufalidis:2012rt, Garoufalidis_2005, Garoufalidis_2008} known as quantum A-polynomials. For example the simplest relation hold for symmetric representations for the trefoil knot:
\begin{equation}
    H^{3_1}_{[r]} + w_{1}(r) \cdot H^{3_1}_{[r-1]} + w_2(r) \cdot H^{3_1}_{[r-2]} = 0
\end{equation}
\begin{align}
\begin{footnotesize}
\begin{aligned}
    w_1(r) &= -\frac{A \left(A q^{2 r}-q^2\right) \left(A q^{2 r}+q^2\right) \left(A^4 q^{8 r}-A^2 q^{4 r+2}-A^2 q^{4 r+6}-A^2 q^{6 r+4}-q^{2 r+8}+q^{4
   r+6}+q^{4 r+8}+q^8\right)}{q^7 \left(q^r-1\right) \left(q^r+1\right) \left(q^3-A q^{2 r}\right) \left(A q^{2 r}+q^3\right)} \\
   w_2(r) &= \frac{A^2 q^{6 r-6} \left(A q^r-q^2\right) \left(A q^r+q^2\right) \left(A q^{2 r}-q\right) \left(A q^{2 r}+q\right)}{\left(q^r-1\right)
   \left(q^r+1\right) \left(q^3-A q^{2 r}\right) \left(A q^{2 r}+q^3\right)}
\end{aligned}
\end{footnotesize}
\end{align}
with "initial conditions" $ H^{3_1}_{[0]} = 1$ and $H^{3_1}_{[1]} = -A^4+A^2 q^2+\frac{A^2}{q^2}$. 

Within the framework of differential expansion \eqref{basicDE} one can study recurrence relations on DE coefficients that known as $\mathcal{C}$-polynomials \cite{Garoufalidis_2006, Cpols}. For the trefoil knot this relation has the following form:
\begin{align}
    \begin{aligned}
       \mathcal{F}^{3_1}_{[s]} + A^{2} q^{2(s-1)} \cdot \mathcal{F}^{3_1}_{[s-1]} = 0 
    \end{aligned}
\end{align}
that is much simpler than the original quantum A-polynomial. The simplicity is not understood completely. One of the possible explanations is that the quantum A-polynomial describes highly nontrivial knot-independent factors of DE \eqref{basicDE}, while $\mathcal{C}$ polynomial deals directly with rather simple DE coefficients.

In recent work \cite{Cpols} was proposed to consider evolution equations on DE coefficients in another direction -- co called $A$-evolution:
\begin{equation}
    \sum_{i=0} t_{i}(A,q) \cdot \mathcal{F}_{[s]}^{\mathcal{K}}(A q^i,q) = 0
\end{equation}
where different terms in the sum undergo a substitution $A \to A q^i$ in the DE coefficients. 

We propose to consider one more evolution direction -- in the knot space . We argue that there is a simple $\mathcal{C}$-polynomial of mixed "representation-knot" evolution for twist knot family. Recall the relation \eqref{mixed C pol} from sec.\ref{GSEtriple}:
\be
(-)^{r} G_{[r],r}^{(n)} = \frac{F_{[r]}^{(n)}}{F_{[r]}^{(1)}}
\ee
that provide a simple form of the resulting formula \eqref{3pretz result}.
Using explicit for of $G_{[r],r}^{(n)}$ functions \eqref{G functions} it takes a form
\begin{equation}
    F^{(n)}_{[r]} =A^{2r} q^{2r(r-1)} \sum_{j=0}^{r} (-1)^{j+1} q^{-j(r-1)} \cdot \frac{[r]!}{[j]! [r-j]!} \cdot F_{[r-j]}^{(n-1)}
\end{equation}
of nontrivial relation that connects DE coefficients for different representations of two twist knots. Remarkably, this relations is nothing but a consequence of formula \eqref{F from KNTZ} involving KNTZ matrix $\mathcal{B}_{i,j} \eqref{KNTZ matr}$:
\begin{equation}
    F_{[r]}^{(n)} = \sum_{j=0}^{i} \ {\cal B}_{r,j} \cdot F_{[j]}^{(n-1)}
\end{equation}
We consider this relation as a $\cal C$-polynomial of {\it mixed} evolution along  representation $[r]$ and twist knot direction $(n)$.
This mixed $\mathcal{C}$-polynomial has much simpler form comparing with pure evolution along representation $[r]$ (see (59) in \cite{Cpols}).
Moreover, this equation has a remarkable property: if one knows all DE coefficients $F_{[r]}$ for a particular twist knot, then one can find DE coefficients for {\it all} other twist knot. In particular,
\begin{equation}
    F^{(-1)}_{[r]} = 1 \hspace{5mm} \Longrightarrow \hspace{5mm} F^{(n)}_{[r]} =  \sum_{j=0}^{r} \ \left( {\cal B}^{n+1} \right)_{r,j}
\end{equation}
that reproduces \eqref{F from KNTZ}.

\section{Conclusion}
 
In this paper we report a remarkably simple formula for the coefficients
of the differential expansion of defect-zero family of antiparallel pretzel knots.
It would be interesting to extend this result in several directions:
\begin{itemize}
    \item \textbf{Non-symmetric representations.}\\
The generalization of the results (33) to higher representations looks straightforward, still technically
involved and interesting. Based of the form of DE coefficients (33) one could try the following ansatz:
 \begin{align}
    G_{R,Q}^{(n)} &:= \sum_{\lambda: \ Q\subset\lambda\subset R}
    V_{R,Q}^{\lambda} \cdot F_\lambda^{(n-1)} \\
    {\cal F}_R^{(n,m,l)} &= \sum_{Q\in R} W_{R,Q}\cdot G^{(n)}_{R,Q} \cdot G^{(m)}_{R,Q} \cdot G^{(l)}_{R,Q}
    \end{align}
 with appropriate functions $W_{R,Q}$ and $V_{R,Q}^{\lambda}$. We expect triple factorization for higher representations for triple pretzels due to the fact that double factorization holds in arbitrary representation for double-braid family. Still it is unclear is this optimistic suggestion will work.
\item \textbf{Other knot families and higher defects.} \\
The formulas of DE coefficients for other pretzels and more general families of knots, including non-
arborescent ones, are less simple -- especially when defect is non-vanishing.
It is interesting to see to what extent the problem will reduce to KNTZ matrix and twist DE coefficients,
like it happened for the zero-defect triple pretzel family.
\end{itemize}
Even more interesting is to find appropriate interpretation of these remarkable formulas for triple pretzels
per se — and the way they deviate from factorization in the double-braid case.
It can happen that the corollaries and applications of such understanding will be no less impressive
than those for the double braids.

\section*{Acknowledgements}
 
We are indebted to E. Lanina, And.Morozov and A.Sleptsov for conversations and help.

This work is supported by the Russian Science Foundation
(Grant No.20-71-10073).

\printbibliography 

\end{document}